\documentclass[prl,aps,twocolumn,showpacs]{revtex4}
\usepackage{graphicx}
\usepackage{amsmath}
\usepackage{bm}
\usepackage{braket}
\usepackage{amstext}
\usepackage{amsxtra}
\usepackage{float}
\usepackage{upgreek}

\usepackage{times}

\newcommand{\kB}{k_{\rm B}}

\newcommand{\unit}[1]{\,\mathrm{#1}}
\newcommand{\bq}{{\mathbf q}}
\newcommand{\bk}{{\mathbf k}}
\newcommand{\gH}{W_{\rm H}}
\newcommand{\tgH}{\tilde{W}_{\rm H}}
\newcommand{\tS}{\tilde{S}}

\begin{document}

\title{Observing Properties of an Interacting Homogeneous Bose--Einstein Condensate: Heisenberg-Limited Momentum Spread, Interaction Energy and Free-Expansion Dynamics}

\author{
Igor Gotlibovych, Tobias F. Schmidutz,  Alexander L. Gaunt, Nir Navon, Robert P. Smith$^\star$,  and Zoran Hadzibabic
}
\affiliation{Cavendish Laboratory, University of Cambridge, J. J. Thomson Avenue, Cambridge CB3 0HE, United Kingdom }

\begin{abstract}
We study the properties of an atomic Bose--Einstein condensate produced in an optical-box potential, using  high-resolution Bragg spectroscopy. For a range of box sizes, up to $70\,\mu$m, we directly observe Heisenberg-limited momentum uncertainty of the condensed atoms. We measure the condensate interaction energy with a precision of $\kB \times 100$~pK and study, both experimentally and numerically,  the dynamics of its free expansion upon release from the box potential. All our measurements are in good agreement with theoretical expectations for a perfectly homogeneous condensate of spatial extent equal to the size of the box, which also establishes the uniformity of our optical-box system on a sub-nK energy scale. 
\end{abstract}

\date{\today}



\maketitle

Ultracold atomic gases produced in harmonic traps are widely used for fundamental studies of many-body quantum mechanics in a flexible experimental setting~\cite{Dalfovo:1999,Pethick:2002,Bloch:2008, Levin:2012}. Recently, it also became possible to produce a  Bose--Einstein condensate (BEC) in an essentially homogeneous atomic gas, held in the quasi-uniform potential of an optical-box trap~\cite{Gaunt:2013}. 
This has opened new possibilities for closer connections with other many-body systems and theories that rely on the translational symmetry of the system (see, e.g.,~\cite{ZinnJustin:1996, Kibble:1976, Zurek:1985, Baym:2001, Andersen:2004RMP, Holzmann:2004, Goldman:2013, Opanchuk:2013, Lu:2013, Sowinski:2013, Radic:2014}).

The first experimental studies of a  Bose gas in a box potential focused on the critical point for condensation and the thermodynamics of the gas close to the critical temperature~\cite{Gaunt:2013, Schmidutz:2014}.
Here, we investigate a box-trapped Bose gas in the low-temperature regime of a quasi-pure condensate. 
While previous experiments~\cite{Gaunt:2013, Schmidutz:2014} established the effective uniformity of the {\it thermal gas} from which the BEC forms, here we directly probe and prove the uniformity of the {\it condensate itself}, which requires measurements on a two orders of magnitude lower energy scale.
We study its coherence, energy, and free expansion from the box trap, employing two-photon Bragg spectroscopy~\cite{Kozuma:1999a, Stenger:1999,Steinhauer:2002,Richard:2003,Ozeri:2005,Papp:2008,Ernst:2010,Lingham:2014} to obtain high resolution measurements of the momentum distribution and interaction energy.
For a wide range of box sizes, extending up to $70\,\mu$m, we directly observe Heisenberg-limited momentum uncertainty of the condensed atoms, corresponding to a fully coherent macroscopic BEC wavefunction spanning the whole box trap. From the interaction shift of the Bragg resonance we deduce the BEC ground-state energy (per atom) with a precision of $\kB \times 100$~pK, and find good agreement with mean-field theory for a perfectly uniform condensate.
Finally, we  study the free time-of-flight (ToF) expansion of a BEC from the box trap. We follow the evolution of the cloud shape and the gradual conversion of the interaction energy into the width of the momentum distribution, and reproduce our observations in numerical simulations based on the Gross-Pitaevskii (GP) equation.

\begin{figure} [b]
\includegraphics[width=\columnwidth]{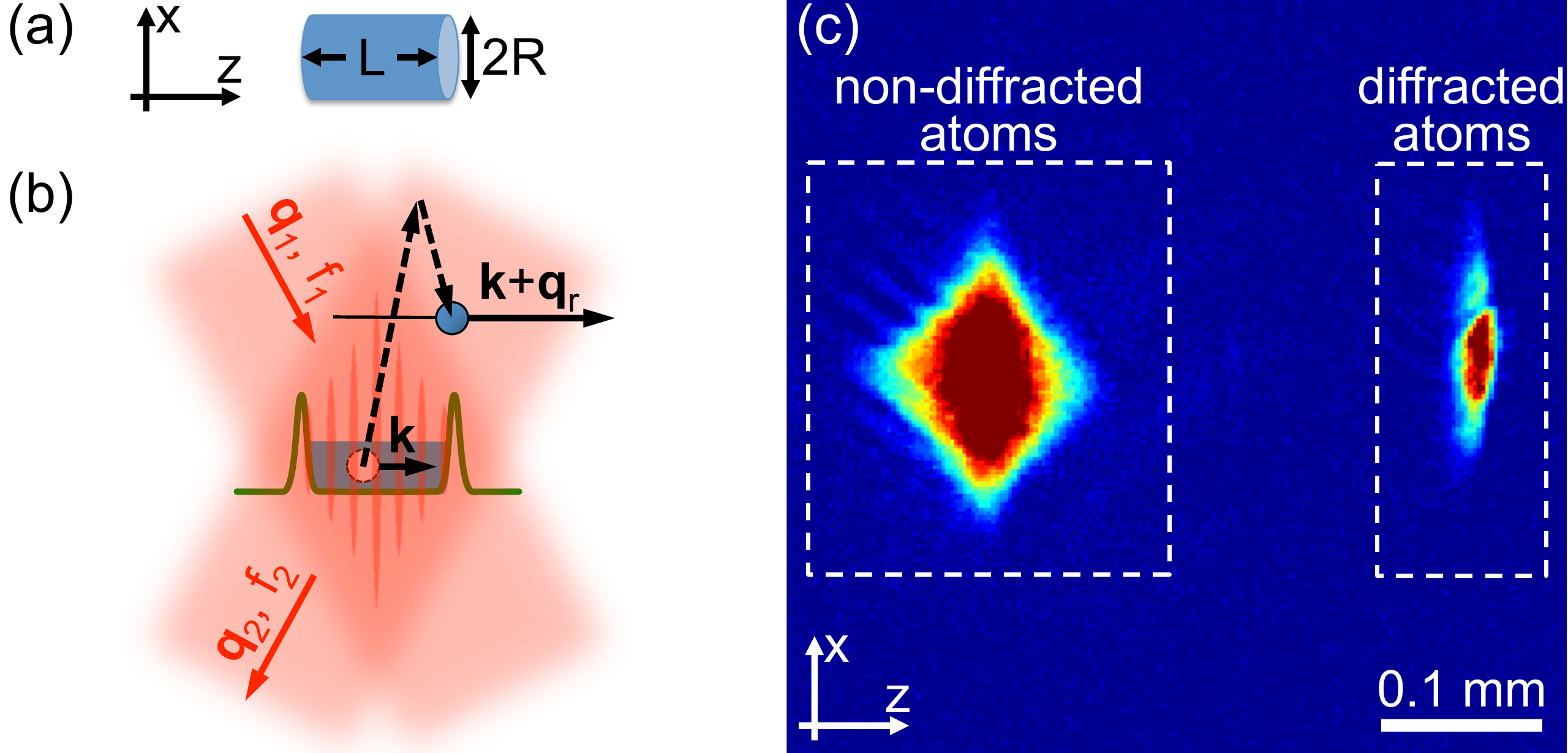}
\caption{(color online) Bragg spectroscopy of a uniform BEC. (a) Shape and orientation of our box trap; in the lab frame the $x$ direction is vertical. (b) A trapped atom with initial momentum $\hbar \bk$ absorbs recoil momentum $\hbar \bq_r = \hbar(\bq_1 - \bq_2)$ aligned with $z$, in a two-photon process that corresponds to Bragg diffraction on a moving optical lattice formed by the interference of the two laser beams. The trap wall height (green) is lower than the energy of the recoiling atoms. 
The Bragg beams and lattice are not drawn to scale; in reality the beams are much larger than the box and the lattice spacing is much smaller. 
(c) Absorption image of the atoms, taken $170$~ms after the start of the Bragg pulse. After the end of the $35$~ms pulse, the non-diffracted atoms are released from the trap.}
\label{fig:cartoon}
\end{figure}

Our apparatus is described in Refs.~\cite{Gotlibovych:2013,Gaunt:2013}. 
We trap $^{87}$Rb atoms in a cylindrical optical box of radius $R \approx 16\,\mu$m and a tuneable length $L = 15-70\,\mu$m [see Fig.~\ref{fig:cartoon}(a)]. Our box is  formed by 532~nm repulsive laser beams, and we use a magnetic field gradient to cancel the linear gravitational potential.
By evaporative cooling in the box trap we produce clouds with condensed fractions $\gtrsim 80\,\%$, at a temperature $T \lesssim 10\,$nK.

In Fig.~\ref{fig:cartoon}(b) we outline our Bragg-spectroscopy setup.
We intersect on the cloud, at an angle of $30^\circ$,  two collimated (3~mm wide) 780~nm laser beams, detuned from the atomic resonance by $6.8$~GHz.
The two beams have wavevectors $\bq_1$ and $\bq_2$, and frequencies $f_1$ and $f_2 < f_1$, respectively. 
An atom that undergoes two-photon stimulated Bragg scattering absorbs energy $hf = h(f_1 - f_2)$ and recoil momentum $\hbar \bq_r = \hbar (\bq_1 - \bq_2)$ aligned with the axial direction of the trap, $z$. The recoil velocity $v_r=\hbar q_r/m\approx 3$~mm/s (where $m$ is the atom mass) is much larger than the spread of velocities in the BEC, $\Delta v_z \sim h/(mL)$, arising due to Heisenberg uncertainty. The diffracted and non-diffracted atoms are thus well separated in velocity space, as qualitatively seen in Fig.~\ref{fig:cartoon}(c).

\begin{figure*} [bt]
\includegraphics[width=\textwidth]{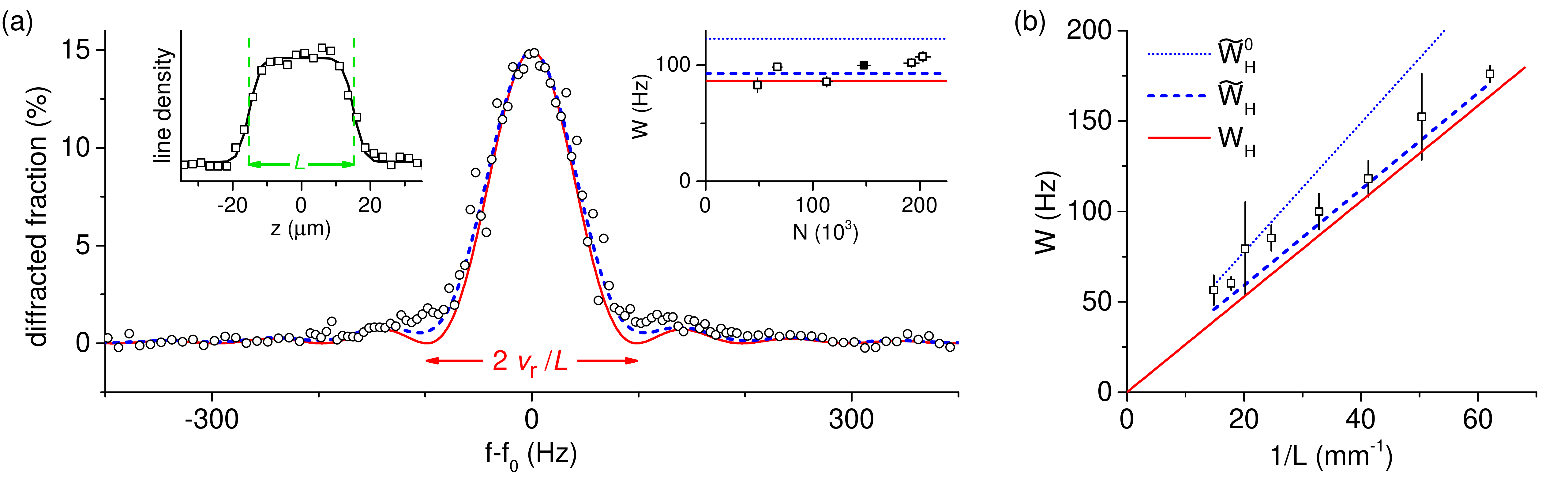}
\caption{(color online) Heisenberg-limited momentum spread in a uniform interacting BEC. 
(a)  Main panel: Bragg spectrum of a trapped BEC of length $L=30\,\mu$m. 
The solid red and dashed blue lines show the theoretical Heisenberg-limited spectra $S(f)$ and $\tS (f)$, respectively;  $S(f)$ is a $\mathrm{sinc}^2$ function obtained by a simple Fourier transform of the top-hat real-space wavefunction, while $\tS (f)$ accounts for the power and duration of the Bragg pulse.
Left inset: $L$ is determined by fitting the in-trap BEC density profile, accounting for the imaging resolution. 
Right inset: Full-width-at-half-maximum of the spectrum, $W$, versus the atom number $N$, for the same $L$. The filled square corresponds to the data in the main panel, with $W = (100\pm3)$~Hz. (All error bars represent $1 \sigma$ uncertainties.) The solid red and dashed blue lines show $\gH = 0.89 \, v_r/L = 87$~Hz and $\tgH=93$~Hz, corresponding to $S(f)$ and $\tS (f)$, respectively. The dotted blue line shows $\tgH^0=123$~Hz, expected for a non-interacting BEC.
(b) $W$ versus inverse box length, $1/L$, showing the expected Heisenberg scaling. Solid red, dashed blue and dotted blue lines show $\gH$, $\tgH$ and $\tgH^0$, respectively.
} 
\label{fig:Heisenberg}
\end{figure*}

Neglecting inter-particle interactions, resonant Bragg scattering of an atom with initial momentum $\hbar \bk$ into the state with momentum $\hbar (\bk + \bq_r)$ occurs for
\begin{equation}
	h f  = h f_r + \frac{\hbar^2}{m}  k_z q_r  \, , 
\label{eq:resonance}
\end{equation}
where $hf_r = \hbar^2 q_r^2/(2 m) \approx h \times  1$~kHz and $k_z q_r= \bk \cdot \bq_r$. Scanning $f$ and counting the number of diffracted atoms [see Fig.~\ref{fig:cartoon}(c)], one can map out the spread of $k_z$ in the cloud. 

In general, interactions in the BEC can modify the width of the Bragg spectrum in two ways: 

First, they modify the {\it momentum distribution}. Repulsive interactions lead to broadening of the wavefunction in real space and thus to narrowing in $\bk$ space~\cite{Stenger:1999}. In a box, the non-interacting ground state along $z$ is $\mathrm{sine}$-like; the corresponding momentum-space wavefunction, obtained by Fourier transform, is $\Psi(k_z) \propto \cos(k_z L/2) / (k_z^2 L^2 - \pi^2)$ and has zero-to-zero width $6\pi/L$. According to Eq.~(\ref{eq:resonance}), the corresponding Bragg spectral line has zero-to-zero width $3v_r/L$ (in frequency units).
On the other hand, the profile of an interacting BEC in the Thomas-Fermi regime is a simple top-hat function of width $L$, neglecting the small edge effects due to the non-zero healing length~\cite{Dalfovo:1999,Pethick:2002}. Our BECs are always deeply in this regime~\cite{footnote:TF}. In this case the momentum-space wavefunction of a fully coherent BEC is $\mathrm{sinc}$-like and the Bragg spectrum is a $\mathrm{sinc}^2$ function, $S(f)$, of zero-to-zero width $2v_r/L$.

Second, interactions can affect the {\it measurement} of the momentum distribution. A condensed atom experiences a mean-field potential $U_0=(4\pi \hbar^2 a/m)n_0$ due to the other ground-state atoms; here $a$ is the s-wave scattering length and $n_0$ the {\it local} BEC density. On the other hand, due to the ``bosonic factor of 2",  a recoiling atom feels an interaction potential $2U_0$~\cite{Pethick:2002}. This shifts the resonant frequency $f$ by $U_0 /h \propto n_0$~\cite{Stenger:1999,Papp:2008,footnote:thermals}, so any spatial variation of $n_0$ leads to inhomogeneous broadening of the Bragg spectrum. 
In our case,  $n_0$ is essentially homogeneous, so this broadening effect is absent and interactions just shift the Bragg resonance. We can thus directly measure the momentum spread in the BEC. 

An important proviso, however, is that to directly observe a Heisenberg-limited momentum spread, the duration of the Bragg pulse must be $\tau \gtrsim L/v_r$. This ensures that the Bragg spectrum is not significantly Fourier-broadened. 
Equivalently, it ensures that a recoiling atom can traverse the box during the pulse, and we thus probe phase coherence across the whole BEC.  
To apply very long Bragg pulses on a trapped cloud, we set the trap depth below the recoil energy $h f_r$, so the diffracted atoms can leave the box without bouncing off the trap walls [see Fig.~\ref{fig:cartoon}(b)].

In Fig.~\ref{fig:Heisenberg} we present our measurements of the momentum uncertainty of the condensed atoms in a box potential. Here $\tau = 35$~ms and the Bragg Rabi frequency was $\Omega_R/(2\pi) \approx 8$~Hz, keeping the fraction of diffracted atoms to $\lesssim 15\,\%$.
We turn off the trap 25~ms after the end of the Bragg pulse, and measure the fraction of diffracted atoms after a further 110~ms of ToF [see Fig.~\ref{fig:cartoon}(c)].

In Fig.~\ref{fig:Heisenberg}(a), we show measurements for $L=30\pm1 \, \mu$m. As shown in the left inset, $L$ is determined by fitting the in-trap BEC density profile with a top-hat function convolved with a Gaussian that accounts for our $5$-$\mu$m imaging resolution~\cite{footnote:SLM}.
 In the main panel we show the measured Bragg spectrum, centred on $f_0 \approx 1$~kHz. 
 The solid red and dashed blue lines show theoretical expectations for a Heisenberg-limited momentum spread. The solid line is the $\mathrm{sinc}^2$ function $S(f)$, while the dashed line is a numerical calculation, $\tS (f)$, which takes into account the small corrections due to the non-infinite $\tau$ and non-zero $\Omega_R$~\cite{IgorPhD}.
The data is clearly extremely close to the Heisenberg limit, corresponding to a fully coherent BEC wavefunction spanning the whole box.
(For corresponding measurements on a harmonically-trapped gas, in momentum and real space, see Refs.~\cite{Stenger:1999} and \cite{Hagley:1999b}, respectively.)

We quantitatively compare different (measured and calculated) Bragg spectra using their full-width-at-half-maximum $W$. In the right inset of Fig.~\ref{fig:Heisenberg}(a), we plot the measured $W$ versus the total atom number $N$, for $L=30\, \mu$m. The solid red and dashed blue lines show the two calculated Heisenberg-limited values, $\gH$ for $S(f)$ and $\tgH$  for $\tS (f)$.
For comparison, we also calculate $\tgH^0$ (dotted blue line) for the $\mathrm{sine}$-like non-interacting ground state of the box potential.
We see that interactions reduce $W$ below $\tgH^0$. Moreover, the measured $W$ shows essentially no dependence on $N$, as expected for a BEC of a spatially uniform density.

In Fig.~\ref{fig:Heisenberg}(b) we summarise our data for a range of box lengths, $L = 15-70\,\mu$m. Plotting $W$ versus $1/L$, we  demonstrate the expected Heisenberg scaling of the momentum uncertainty.

We now turn to the study of the ground-state energy of a uniform interacting BEC, which is seen in the {\it shift}  of the Bragg resonance, $f_0$, from the recoil frequency $f_r$. 
Thanks to the unprecedented narrowness of our Bragg spectra, we measure such shifts with a precision of $2$~Hz, corresponding to an energy of $\kB \times 100$~pK. 

\begin{figure} [t]
\includegraphics[width=\columnwidth]{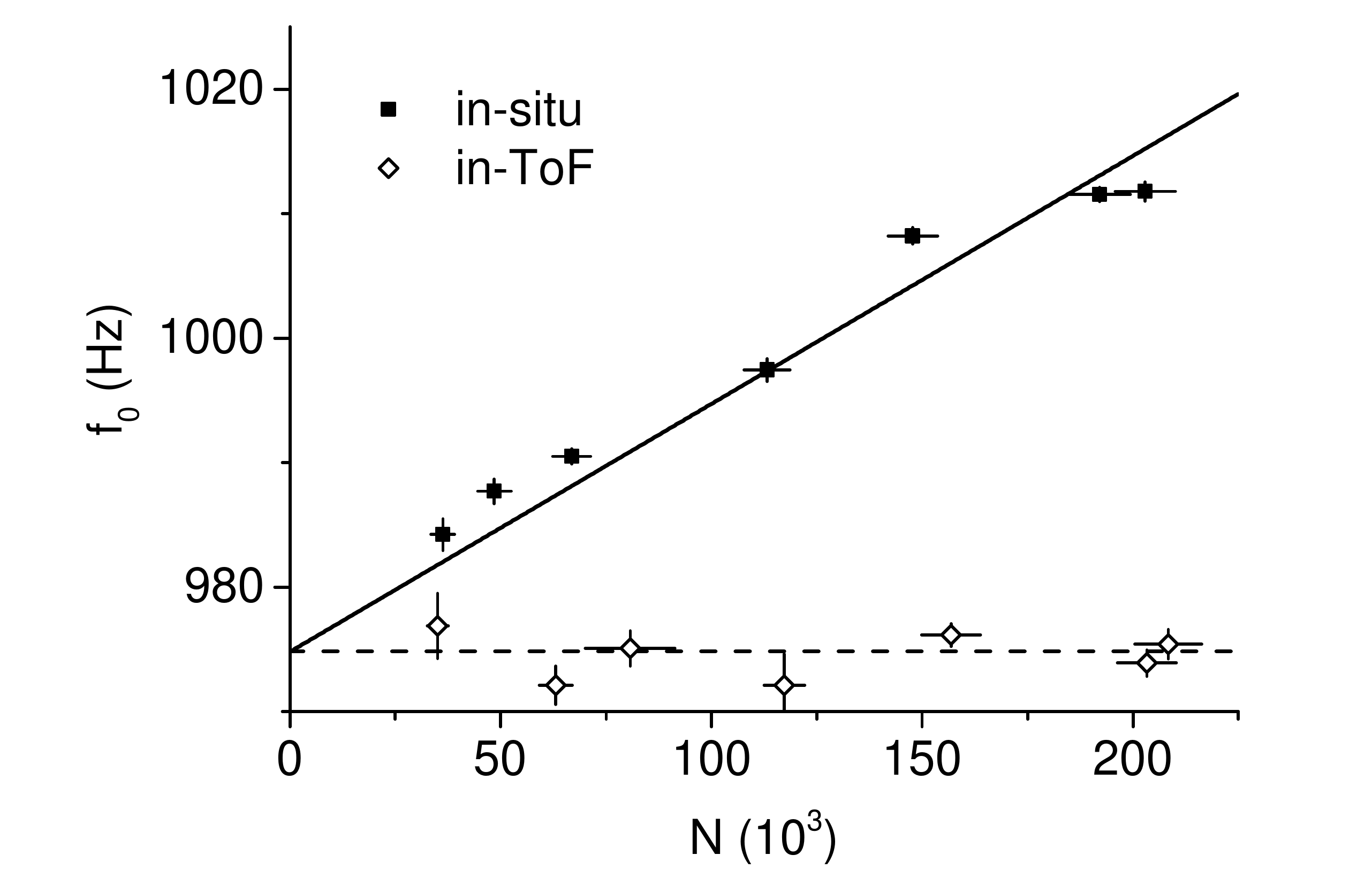}
\caption{Interaction energy in a uniform BEC, for $L=30\,\mu$m. We plot $f_0$ versus $N$ for Bragg pulses applied in-situ (solid squares) and after $50$~ms of ToF (open diamonds). Dashed line: $f_0 \approx f_r \approx 975\unit{Hz}$, solid line: $f_0= f_r + \alpha N$, with $\alpha \approx 20 \times 10^{-5}$~Hz/atom.}
\label{fig:Nf0}
\end{figure}

In Fig.~\ref{fig:Nf0} we plot $f_0$ versus $N$, for a fixed $L = 30\,\mu$m and for two sets of Bragg spectra:  (i) the ``in-situ" spectra (solid symbols) are taken as above, with the main cloud trapped during the Bragg pulse, and (ii) the ``in-ToF" spectra (open symbols) are taken by releasing the BEC from the box and letting it expand for 50~ms before applying the pulse. After long ToF the atomic density is negligible and the $N$-independent $f_0$ provides a good measurement of $f_r \approx 975$~Hz.

For the in-situ spectra, within mean-field theory and for an infinitesimal Bragg transfer ($\Omega_R \tau \rightarrow 0$), we expect:  
\begin{equation}
	\Delta f_0 \equiv f_0 - f_r  = \alpha N  , 
\label{eq:shift}
\end{equation}
with $\alpha = 2 \eta \hbar a/(mV)$, where $\eta$ is the condensed fraction~\cite{footnote:thermals} and $V$ is the volume of the box. 
From in-situ images we get $V=(25 \pm 2) \times 10^3 \,\mu$m$^3$.
We assess $\eta = 0.8 \pm 0.1$ from the maximal fraction of Bragg-diffracted atoms [$15\%$ in Fig.~\ref{fig:Heisenberg}(a)]; here the uncertainty indicates variations between experimental runs. This estimate is further supported by ``BEC filtering" introduced in Ref.~\cite{Gerbier:2004c}, i.e., using a short (4~ms) Bragg $\pi$-pulse to separate the BEC from the residual thermal component in ToF~\cite{footnote:eta}.
We thus obtain a theoretical prediction $\alpha_{\rm th} = (24 \pm 2) \times 10^{-5}$~Hz/atom.

From a linear fit to the in-situ data (solid line in Fig.~\ref{fig:Nf0}) we get $\alpha_{\rm exp} = (20 \pm 1) \times 10^{-5}$~Hz/atom, slightly below $\alpha_{\rm th}$. This small difference can be attributed to the $15\%$ depletion of the ground-state population during the pulse. For our largest $N$ we took additional measurements with a reduced $\Omega_R$~\cite{TobyPhD} and extrapolating to $\Omega_R =0$ we get a slightly revised $\tilde{\alpha}_{\rm exp}=(23 \pm 1) \times 10^{-5}$~Hz/atom~\cite{footnote:systematics}.

\begin{figure} [t]
\includegraphics[width=\columnwidth]{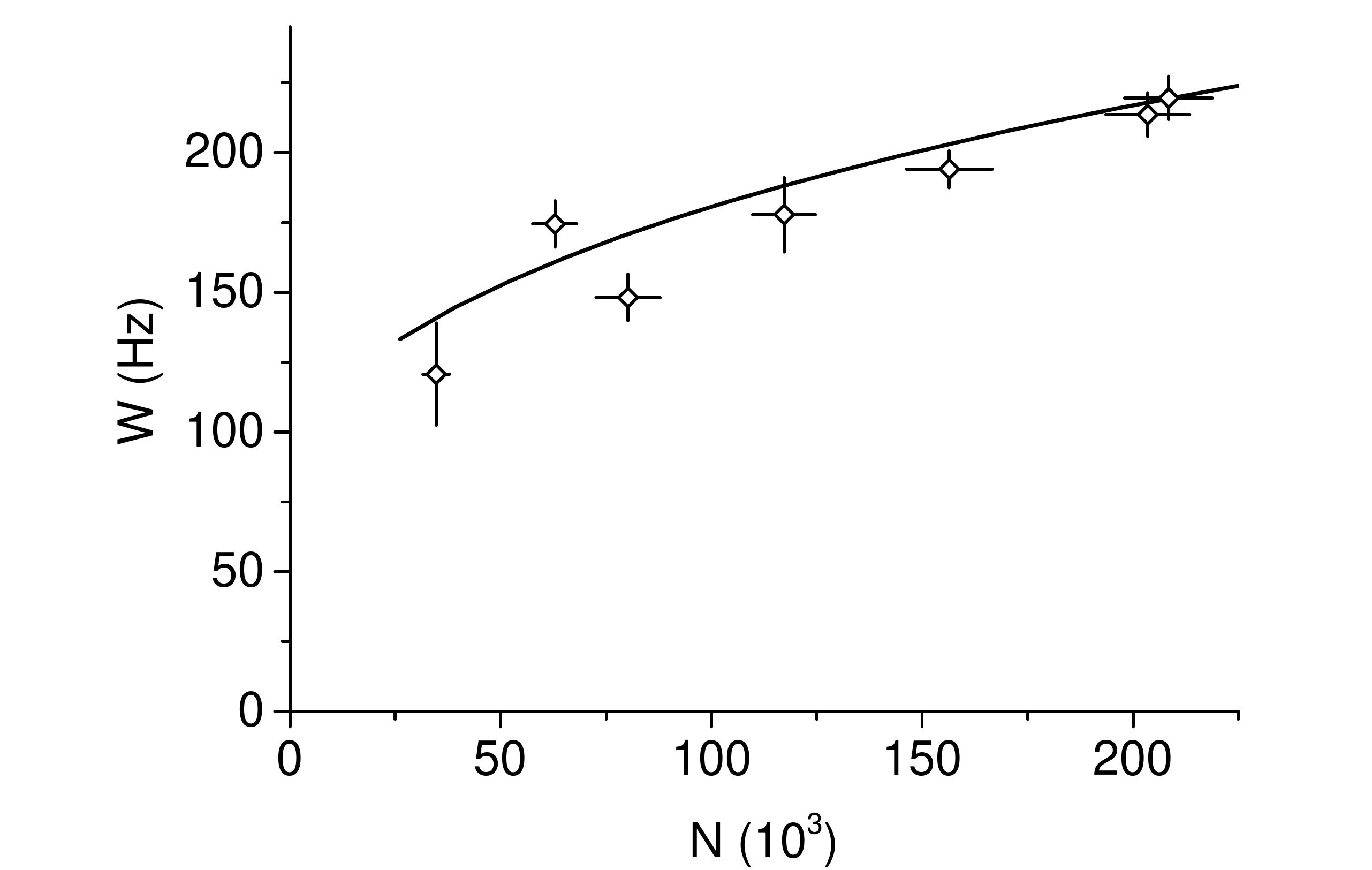}
\caption{Spectral width for Bragg pulses applied after 50~ms of ToF expansion, for $L=30\,\mu$m. The $W$ values are extracted from the same in-ToF spectra as the $f_0$ values in Fig.~\ref{fig:Nf0}. The solid line shows a numerical simulation based on the GP equation.}
\label{fig:N_FWHM}
\end{figure}

Complementary to Fig.~\ref{fig:Nf0}, in Fig.~\ref{fig:N_FWHM} we plot $W$ values for the in-ToF Bragg spectra. Qualitatively, $W$ now grows with $N$ because during ToF interaction energy gets converted into kinetic energy. Quantitatively, the problem of the expansion of an interacting BEC from a box potential has not been solved analytically (see ~\cite{Castin:1996} for the harmonic-trap case). However, we find good agreement between our data and numerical simulations based on the GP equation (solid line). In our simulations, we neglect the small thermal component and use the measured in-trap BEC energy ($\propto \tilde{\alpha}_{\rm exp} N$) to predict $W$ in ToF~\cite{footnote:harmonic}.

\begin{figure} [tb]
\includegraphics[width=\columnwidth]{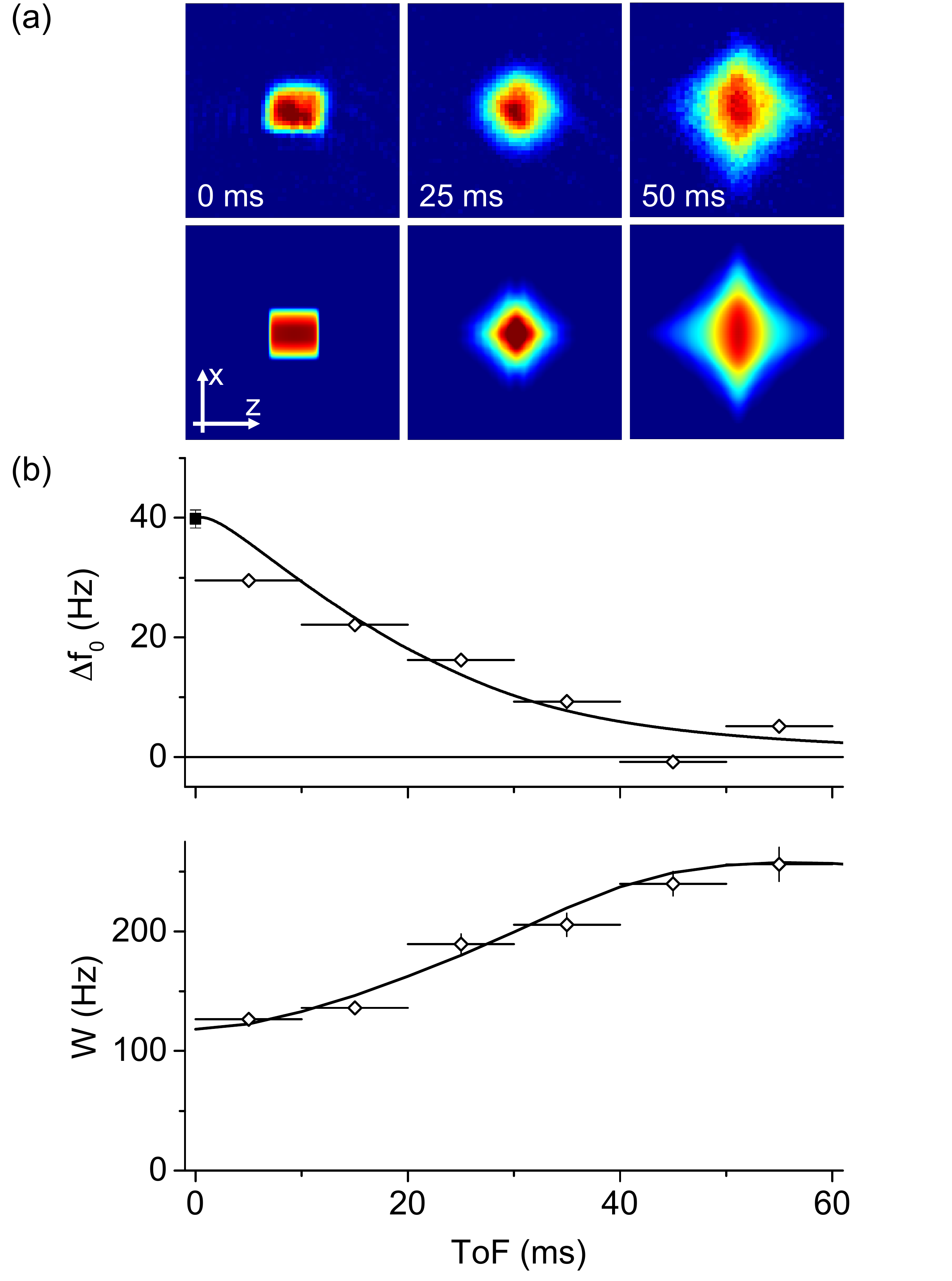}
\caption{(color online) Free expansion of a BEC released from the box trap. Here, $L = 30\,\mu$m and $N \approx 200 \times 10^3$. (a) Absorption images (top) and simulations (bottom) of the atomic distributions. (b) Evolution of  $\Delta f_0$ and $W$ during ToF.  The solid lines show numerical simulations based on the GP equation. The solid square corresponds to the in-trap data from Fig.~\ref{fig:Nf0}. The horizontal error bars indicate our temporal resolution, limited by the Bragg-pulse duration.}
\label{fig:movie}
\end{figure}

Finally, we study the evolution of the BEC in ToF, for $L = 30\,\mu$m and $N \approx 200 \times 10^3$.
In Fig.~\ref{fig:movie}(a) we show absorption images of the expanding cloud (top), and the corresponding GP-based simulations (bottom).
The simulations reproduce the characteristic diamond shape that emerges during ToF, also seen in Fig.~\ref{fig:cartoon}(c).
Qualitatively, this conversion to a diamond shape is the analogue of the inversion of the aspect ratio of a BEC released from an anisotropic harmonic trap. In both cases the force driving the initially accelerating expansion is given by the gradient of the atomic density (i.e., the gradient of the interaction-energy density) and in both cases the fastest moving wavefronts develop at the points where the curvature of the constant-density surfaces is minimal during the early stages of the expansion.

In Fig.~\ref{fig:movie}(b) we show the gradual decay of $\Delta f_0$ and growth of $W$ for in-ToF Bragg spectra, again finding good agreement with our simulations (solid lines). Note that in these experiments, and simulations, we reduced $\tau$ to 10~ms and increased $\Omega_R/(2\pi)$ to 28~Hz, reducing our spectral resolution in order to improve the time resolution.

In conclusion, we have characterised the ground-state properties of an interacting homogeneous Bose gas, including the Heisenberg-limited momentum distribution, the interaction energy, and the free-expansion dynamics.
An important by-result of our measurements is that they place the most stringent bound so far on the spatial uniformity of an ultracold gas produced in our optical box. 
While earlier (thermodynamic) studies established uniformity on a $30-100$~nK energy scale~\cite{Gaunt:2013,Schmidutz:2014}, all our present measurements indicate that our gas behaves as a homogeneous system down to a sub-nK energy scale (corresponding to 20~Hz in frequency units). Such a high degree of uniformity offers great promise for future studies of correlation physics in a homogeneous gas in the $T \rightarrow 0$ limit, for example for the preparation and detection of topologically protected states~\cite{Goldman:2013}.

We thank Sadi Ayhan for experimental assistance, Nir Davidson for useful discussions, and Richard Fletcher and Martin Robert de Saint Vincent for comments on the manuscript.
This work was supported by EPSRC (Grant No. EP/K003615/1), AFOSR, ARO and DARPA OLE. N.N. acknowledges support from Trinity College, Cambridge. R.P.S. acknowledges support from the Royal Society.  I.G. and T.F.S. contributed equally to this work.


\begin{thebibliography}{10}
\bibitem{footnote:TF}
For our box dimensions, the Thomas-Fermi regime is reached for a few thousand condensed atoms. For all the measurements presented here the BEC atom number is $> 30 \times 10^3$.
\bibitem{footnote:thermals}
Both before and after undergoing Bragg scattering an atom initially in the BEC feels the same ``double" interaction with any thermal component of the cloud. Bragg spectroscopy is thus sensitive only to the ground-state interactions.
Once a recoiling atom leaves the box its interaction energy changes; however, the atom is no longer Bragg-coupled to the stationary BEC, so this does not affect the spectroscopic frequency shift.
\bibitem{footnote:SLM}
Within experimental errors, the measured $L$ is consistent with the value we calculate from the settings of the spatial light modulator used to create and align the trapping beams~\cite{Gaunt:2013}.
\bibitem{footnote:eta}
Note that the small thermal fraction is not directly measurable from either standard ToF images or the Bragg spectra such as shown in Fig.~\ref{fig:Heisenberg}(a).
\bibitem{footnote:systematics}
The systematic uncertainty in $\tilde{\alpha}_{\rm exp}$, due to the systematic uncertainty in our absolute atom-number calibration, is 10\%.
\end{thebibliography}

\end{document}